\documentclass[12pt,preprint]{aastex}

\slugcomment{Submitted to ApJ Letters}

\shorttitle{Estimating the birth period of pulsars}
\shortauthors{de Jager}

\begin{document}


\title{Estimating the birth period of pulsars through \\
GLAST/LAT observations of their wind nebulae}


\author{O.C. de Jager\altaffilmark{1}}
\affil{Unit for Space Physics, North-West University, Potchefstroom 2520, South Africa}
\altaffiltext{1}{South African Department of Science \& Technology and National Research Foundation
Research Chair: Astrophysics \& Space Science}



\begin{abstract}
In this paper we show that the high energy $\gamma$-ray flux in the GeV domain from mature
pulsar wind nebulae (PWN) scales as the change in rotational kinetic energy $I(\Omega_0^2-\Omega^2)/2$ since birth, rather than the
present day spindown power $I\Omega\dot{\Omega}$. This finding holds as long as the lifetime of inverse Compton
emitting electrons exceeds the age of the system. For a typical $\gamma^{-2}$ electron spectrum,
the predicted flux depends mostly on the pulsar birth period, conversion efficiency 
of spindown power to relativistic electrons and distance to the PWN, so that first order estimates of the birth
period can be assessed from {\it GLAST/LAT} observations of PWN. For this purpose we derive an analytical expression.
The associated (``uncooled'') photon
spectral index in the GeV domain is expected to cluster around $\sim 1.5$, which is bounded at low energies
by an intrinsic spectral break, and at higher energies by 
a second spectral break where the photon index steepens to $\sim 2$ due to radiation losses.
However, if the spectral parameters deviate from the abovementioned assumptions, we can 
combine {\it GLAST/LAT} observations with multiwavelength data from radio to TeV $\gamma$-ray observations
to improve such birth period estimates. From the {\it ATNF} pulsar catalog 
we expect a total of 50 galactic PWN to be detectable if the ratio of birth period to current period is $<0.5$,
but 30 for a ratio $<0.75$.
Mature PWN are expected to have expanded to sizes larger than currently known PWN, resulting in relatively
low magnetic energy densities and hence survival of GeV inverse Compton emitting electrons. Whereas such a PWN may be radio and X-ray quiet
in synchrotron radiation, it may still be detectable as a {\it GLAST/LAT} source as a result of the relic electrons in the PWN.
\end{abstract}





\section{Introduction}

The most reliable method of determining the initial spin period of a pulsar
is to have a historical association of the supernova explosion, combined
with a measurement of the pulsar braking index: For Crab, \citet{lpgs93}
estimated a birth period of $P_0=19$ ms, whereas a value of $P_0=63$ ms was
determined for \objectname{PSR B1509-58}. The pulsar current technique was
introduced by \citet{pb81} and \citet{vn81}, but suffers from small number
statistics \citep{l93}. \citet{n87} and \citet{ec89} proposed a technique
based on the pulsar luminosity function technique, injection and the presence
of pulsar wind nebulae (PWN), so that injection at $P_0\sim $ few hundred milliseconds
would result in the absence of a PWN. \citet{vw01} have shown that the pulsar wind nebula
radius is a function of the birth period of a pulsar, and that the ratio of this radius
relative to the supernova remnant forward shock radius provides a measure of the birth period as well.

In this paper we show that high energy gamma-ray observations of pulsar wind nebulae in the GeV domain
can also give us a measurent of the birth period of the associated pulsar wind nebula. The reasoning
behind this is simple: the relatively low energy leptons radiating GeV $\gamma$-rays via inverse Compton
scattering on the cosmic microwave background (CMBR) and the galactic radiation fields
do not suffer from radiation losses as e.g. X-ray synchrotron emitting electrons do. In this case 
the GeV flux should be proportional to the total energy in leptons ejected since birth, which
in turn is a fraction $\eta$ of the total rotational kinetic energy $E_{\rm rot}=I(\Omega_0^2-\Omega^2)/2$ released since birth.
Here $\Omega_0=2\pi/P_0$ and $\Omega$ are respectively the birth and current pulsar angular frequencies.


Normally such a technique would suffer from uncertainties in
PWN parameters, but we will show that if we assume a particle injection spectral index of 2 above 
the radio domain, the expected GeV flux can be expressed as a simple analytic formula, in which case
the uncertainty in $P_0$ scales linearly with the uncertainty in the distance, but less sensitively on other less known parameters
such as $\eta$, the maximum particle energy $\gamma_{\rm max}m_ec^2$ and the intrinsic spectral break energy $\gamma_bm_ec^2$ above the radio domain. The latter typically divides the injection spectrum into a $\sim \gamma^{-2}$
component above this break, but a much harder spectrum at $\gamma \ll \gamma_b$ to account for the flat spectra seen from radio PWN.
The requirement for such an intrinsic break in the \objectname{Crab nebula} and other PWN was discussed respectively by \citet{aa96} and \citet{c04}.

Whereas such an analytical expression helps us with population studies and to get approximate estimates for $P_0$, 
a {\it GLAST/LAT} detection with a clear multiwavelength PWN association should not rely on this expression to
obtain an accurate estimate of $P_0$: The multiwavelength synchro-Compton spectrum should give us more information 
about particle spectra, break energies and conversion efficiencies through proper numerical modeling, 
although the principle remains the same as expressed below.

\section{The electron spectrum contributing to the {\it GLAST/LAT} spectral range}
We adopt the injection spectrum of \citet{vd06} for electrons at the pulsar wind shock
\begin{equation}
Q(\gamma,t)=\left(
\begin{array}{l}
Q_0(t)(\gamma/\gamma_b)^{-p_1}\,\,{\rm for}\,\,\gamma<\gamma_b\\
Q_0(t)(\gamma/\gamma_b)^{-p_2}\,\,{\rm for}\,\,\gamma_b<\gamma<\gamma_{\rm max}
\end{array}\right),
\end{equation}
with $\gamma_bm_ec^2$  the intrinsic break energy. For $p_1\sim 1$ we retrieve the typical flat radio
spectra of PWN where $F_\nu\propto \nu^{-\alpha}$ with $\alpha\sim 0$. The high energy spectrum is steeper
and can extend up to ultra high energies ($\gamma_{\rm max}m_ec^2\sim 10^{14}$ eV to $>10^{15}$ eV). The corresponding spectral index 
$p_2\sim 2$ as a result of particle acceleration at or near the pulsar wind shock. This is supported
by observations at X-ray and very high energy $\gamma$-ray energies: 
At the pulsar wind shock (radius $R_s$) we observe X-ray and VHE $\gamma$-ray photon indices 
$\Gamma_s=(p_2+1)/2\sim 1.5$, whereas
for $r\gg R_s$ the electron spectrum cools due to synchrotron losses ($p_2$ increases by unity), 
resulting in a value of $\Gamma_x=(p_2+2)/2\sim 2$ as illustrated by \citet{d08}, who combined the results from
several PWN in the form of a plot of $\Gamma_x$ vs $\Gamma_s$. This reveals a cluster near $\Gamma_s\sim 1.5$,
evolving to $\Gamma_x\sim 2$. In this paper we will assume $p_2=2$, resulting in a logarithmic suppression
of uncertainties in $\gamma_b$ and $\gamma_{\rm max}$, which is important for a relative accurate determination of the 
birth period. Note however that there are deviations: There are a few cases for which $p_2\sim 3$, in which case 
a multiwavelength/numerical modelling approach is called for, as discussed in the Introduction.

In general we find a critical energy $\gamma_{\rm rad}m_ec^2$, such that radiation losses are unimportant
for $\gamma \ll \gamma_{\rm rad}$ (the ``uncooled'' domain) and $R_s<r<R_{\rm PWN}$ (with $R_{\rm PWN}$ the PWN radius), 
in which case the total observed spectrum 
is reflected by the injection indices $p_1$ and $p_2$. For $\gamma \gg \gamma_{\rm rad}$ (the ``cooled'' domain),
the corresponding electron spectral index steepens to $p_2+1$ due to radiative cooling, in which case
total nebular photon spectra of the form $E_{\gamma}^{-2}$ are observed.  
An estimate of $\gamma_{\rm rad}$
can be obtained by setting the timescale for radiative losses (assuming isotropic lepton pitch angles)
equal to the age $\tau_{\rm age}$ of the system
and rewriting this electron energy in terms of its inverse Compton scattered $\gamma$-ray energy giving
(assuming scattering on the CMBR):
\begin{equation}
E_{\gamma}(\rm rad)= \frac{4}{3}(2.7kT)\gamma_{\rm rad}^2\sim (0.50\,{\rm TeV})\left(\frac{B}{10\,\mu G}\right)^{-4}\left(\frac{\tau_{\rm age}}{10\,{\rm kyr}}\right)^{-2}.
\label{ebreak}
\end{equation}
For $B<3\mu$G this expression must be corrected for inverse Compton losses on the CMBR.
The spectral breaks of the two mature {\it H.E.S.S.} sources \objectname{HESS J1825-137} 
($E_{\gamma}(\rm rad)\sim 2$ TeV; $\tau_{\rm age}=20$ kyr -- \cite{Aha06b}) and the cocoon of \objectname{Vela X} 
($E_{\gamma}(\rm rad)\sim 13$ TeV; $\tau_{\rm age}=11$ kyr -- \cite{Aha06a}) can be explained with field strengths between 4 and $5\mu$G.
Furthermore, the relatively faint X-ray synchrotron flux from these sources support such low field strengths
(see also \cite{d08}). The newly detected PWN \objectname{HESS J1718-385} also shows a spectral break at 7 TeV \citep{Aha07}
and the discovery of an X-ray PWN around the associated pulsar \objectname{PSR J1718-3825} implies a field strength
$<5\mu$G \citep{Hin07}. For a characteristic age of 90 kyr the predicted break energy is however in the multiple GeV range.
If the birth period is not much less than the current 75 ms period, it would be possible to reproduce the break \citep{Hin07}, but
for this we need to know the pulsar braking index.

Assuming evolved PWN with radiation breaks not lower than 10 GeV and $B\sim 5\mu$G, {\it GLAST/LAT} should be able to detect
PWN between the intrinsic break energy and $\sim 10$ GeV with a (``uncooled'') photon index near 1.5 if the age
\begin{equation}
\tau_{\rm age}<(280\,{\rm kyr})\left(\frac{B}{5\,\mu{\rm G}}\right)^{-2}.
\label{10gev}
\end{equation}
Having set the constraints for detecting the specified injection spectral indices, we can now proceed
with the high energy flux calculations: Following \citet{sd03} and \citet{vd06}, the energy
equation can be written in terms of the present day spindown power $L(t)$ giving
\begin{equation}
m_ec^2\int Q(\gamma,t)\gamma d{\gamma}=\eta L(t).
\end{equation}
From this expression the normalisation constant $Q_0(t)$ for $p_2=2$ follows:
\begin{equation}
Q_0(t)=\frac{\eta L(t)}{m_ec^2\gamma_b^2a} \;\;{\rm with}\,\,a=\left(\frac{1}{2-p_1}+ln\frac{\gamma_{\rm max}}{\gamma_b}\right),
\end{equation}
and $\eta$ the conversion efficiency of spindown power into energetic particles. 
Since we observe at electron energies well below $\gamma_{\rm rad}$ (assuming escape losses
are also unimportant), the total nebular particle spectrum at $r>R_s$ can be derived from the 
transport equation to give
\begin{equation}
\frac{dN(\gamma,T_{\rm age})}{d\gamma}=\int_0^{T_{\rm age}}Q(\gamma,t)dt\sim \frac{\overline{\eta}\Delta {\rm KE}_{\rm rot}}{am_ec^2}\gamma^{-2}.
\end{equation}
The time integral $\int L(t)dt$ up to the present age is given by the change in rotational kinetic energy 
$\Delta{\rm KE}_{\rm rot}$ $=I(\Omega_0^2-\Omega^2)/2$ $=2\pi^2I(1/P_0^2-1/P^2)$.

\section{The expected $\gamma$-ray flux and the sensitivity of {\it LAT} to birth periods}
The differential gamma-ray spectrum resulting from the inverse Compton scattering of the CMBR (with $T=2.7$K) is given by \citet{bg70}.
We then replace the electron spectral normalisation constant with the abovementioned rotational kinetic energy to give a 
total differential photon rate of
\begin{equation}
\frac{dN_\gamma}{dtdE_\gamma}=\frac{r_0^2}{\pi\hbar^3m_ec^4}\frac{\overline{\eta}}{a}
\Delta {\rm KE}_{\rm rot}(kT)^{7/2}F(p_2)E_\gamma^{-1.5}.
\end{equation}
Note the photon spectral index of 1.5 resulting from electrons radiating with a $p_2=2$ spectral index. The function $F(p_2)=5.3$ for $p_2=2$.
\citet{d95} introduced the necessity of including the contribution of far infrared (FIR) photons from galactic dust grains (at a temperature of $\sim 25$K) to the inverse Compton process for PWN for which external Compton scattering is more important than
the synchrotron-self-Compton process. 
In general, if $U_i$ (in units of eV/cm$^{3}$) is the energy density of an ambient target photon field with corresponding temperature $T_i$, the additional flux in the Thomson limit due to species $i$ would then be given by (in the energy range where $p_2=2$ contribute to the $\gamma$-ray flux - from \cite{bg70})
\begin{equation}
\frac{dN_\gamma(T_i)}{dtdE_\gamma}\sim 6.5T_i^{-1/2}U_i\frac{dN_\gamma(T=2.7\,K)}{dtdE_\gamma}.
\end{equation}
With this derivation we assumed that the energy density at temperature $T_i$ is a constant times that of a pure
black body at the same temperature. 
The energy density of this FIR is $U\sim 1$ eV/cm$^{-3}$ within a galactocentric radius of $\sim 4$ kpc, 
decreasing to $\sim 0.3$ eV/cm$^{-3}$ towards the solar neighbourhood \citep{pms06}.
From this expression for the $\gamma$-ray flux it is clear that high temperature photon fields do not contribute, unless
the corresponding photon energy density is relatively large. We will therefore only add the contributions for the CMBR and FIR dust fields.

By selecting $p_2=2$, the constant $a$ in $Q_0(t)$ can vary between 6 and 12 for $p_1\sim 1$, $\gamma_{\rm max}mc^2=10^{14}$ eV (Vela-like) 
to $2\times 10^{15}$ eV (Crab-like) and a break energy of $\gamma_bm_ec^2$ between $10^{11}$ eV and $3\times 10^{12}$ eV
to reproduce the observed break frequencies near $10^{13}$ Hz and $10^{16}$ Hz respectively in the Crab nebula.
We will normalise $a$ to a value of 10 by setting $a=10a_{10}$. In general, since the birth period scales as $a^{-1/2}$, 
uncertainties in the abovementioned parameters are significantly suppressed since we take the square root of a logarithm.  

We can now write the total high energy $\gamma$-ray spectrum from a PWN in the {\it GLAST/LAT} GeV domain 
(i.e. where $\gamma_b<\gamma<\gamma_{\rm rad}$), where the electron spectral index over the whole nebula
remains $\sim 2$, so that the photon index is $\sim 1.5$. If the pulsar is also sufficiently spun down ($P_0 \ll P$),
we can neglect the $P$-term in $\Delta {\rm KE}_{\rm rot}$, so that
\begin{equation}
\frac{dN_\gamma}{dtdE_\gamma}=4\times 10^{-8}\frac{(1+1.3U_{25})}{a_{10}P_{40}^2d_{\rm kpc}^2}I_{45}\overline{\eta} E_{\gamma}^{-1.5}
\label{spectrum}
\end{equation}
in units of cm$^{-2}$s$^{-1}$GeV$^{-1}$. Here $I_{45}$ is the moment of inertia normalised to $10^{45}$ g.cm$^2$, whereas the birth
period is normalised to 40 ms: $P_0=(40\,{\rm ms})P_{40}$, following the finding of \citet{vw01} that many pulsars in composite
remnants were born with $P_0\sim 40$ ms. From the integral flux above 1 GeV we can calculate the {\it LAT} sensitivity
to detect a given birth period: For a 50 hr, $5\sigma$ flux sensitivity of $2\times 10^{-10}$ cm$^{-2}$s$^{-1}$ above 1 GeV 
it is clear that {\it LAT} should be sensitive to birth periods in the range
\begin{equation}
P_0<(44\,{\rm ms})(1+1.3U_{25})^{1/2}\left(\frac{\overline{\eta}}{0.3}\frac{I_{45}}{a_{10}}\right)^{1/2}\left(\frac{d}{10\,{\rm kpc}}\right)^{-1}.
\end{equation}
For example, for a source at a distance of 10 kpc, but within a galactocentric radius of 4 kpc where $U_{25}\sim 1$ eV/cm$^{-3}$,
GLAST should be able to detect birth periods less than 67 ms, whereas for a source at a distance of 2 kpc ($U_{25}\sim 0.3$ eV/cm$^{-3}$)
birth periods as long as 260 ms can be detected, assuming a realistic conversion efficiency of $\overline{\eta}=0.3$.

Assuming birth periods $P_0\le \epsilon P$ (with $\epsilon<1$), we can use this flux prediction (including the $P$-term)
to determine the number of pulsars with expected GeV fluxes above the {\it GLAST/LAT} sensitivty threshold.
Using the {\it ATNF} pulsar catalog of \citet{m05}, we select all pulsars with characteristic ages less than 280 kyr (according
to the condition set by Eqn. (\ref{10gev})) and for each of these pulsars we calculate the integral flux above 1 GeV
from Eqn (\ref{spectrum}) for a given period and selected value of $\epsilon$ (the current period was included for this
calculation). Note that we only take the total number from this catalog and did not correct for
selection effects such as dispersion measure constraints and off-beam pulsars. Adding such potentially unseen pulsars
should increase the numbers presented below.
The result from the existing catalog is 
a total of 50 detectable pulsars for $\epsilon=0.5$, but only 30 for
$\epsilon=0.75$. Note however that the true birth periods are not expected to scale as a fixed fraction $\epsilon$ of
the current period -- the only purpose of these estimates is to show that a relatively large number of PWN are
expected to be visible even if $P_0$ differ by only 25\% from $P$.



\section{Discussion and conclusion}
In this paper we derived the expected GeV $\gamma$-ray spectrum of PWN under the assumption
that the particle spectrum in the uncooled domain ($\gamma<\gamma_{\rm rad}$)
is $p_2\sim 2$, as observed from the injection spectra at most PWN shocks. When
observing mature PWN (i.e. those for which the age is significantly larger than the spindown
timescale), we arrive at a fresh new perspective w.r.t. to the different
multiwavelength domains:

For the GLAST/LAT domain above 1 GeV, the observed $\gamma$-ray luminosity would
scale as the change in rotational kinetic energy $I(\Omega_0^2-\Omega^2)/2$, rather than
the present day spindown power $I\Omega\dot{\Omega}$. This is because the
electrons contributing to the {\it LAT} energy domain survive since the
earliest epoch when the spin period was close to the birth period. 
We therefore predict that the GeV luminosity and $I\Omega\dot{\Omega}$
would be uncorrelated. There is however a caviat to this argument:
The field strength during earlier epochs is expected to be larger, so that
GeV emitting electrons may have been burnt off during an earlier epoch of high $B$. 
However, for the PWN of PSR\,B1509-58 the nebular field strength is 
between $8\mu$G \citep{dup95,gae02} and $17\mu$G \citep{Aha05}, given an age of $\sim 1,700$ years, whereas
Eqn (\ref{ebreak}) requires a field strength of $B<64\mu$G after 1,700 years to force a radiation spectral
break above 10 GeV. It is thus clear that this requirement is easily achieved for this example. The same can be
shown for other relatively young VHE emitting PWN, such as \objectname{Kes 75}, \objectname{G21.5-0.9} and \objectname{G0.9+0.1},
except for the \objectname{Crab Nebula} for which $B\sim 2\times 10^{-4}$ G after 950 years. In general, 
the expansion history of the PWN, given the environment into which the associated supernova remnant forward shock
expands, is expected to determine the time evolution of PWN field strength. See also \cite{Rey84}.

For the very high energy $\gamma$-ray (ground based) domain, we detect
inverse Compton radiation from electrons around the radiation maximum  $\gamma_{\rm rad}$
as reviewed by \citet{d08} for PWN older than 10 kyr (see the discussion on
\objectname{Vela X} and \objectname{HESS J1825-137}). This is possibly also the case for HESS\,J1718-385 \citep{Aha07,Hin07}.
Below this maximum the photon spectral index is near 1.5 due to relatively old electrons
ejected since birth. This hard component should extend into the {\it GLAST/LAT}
range as discussed in this paper. However, electrons with energies $\gamma \gg \gamma_{\rm rad}$
have lifetimes shorter than the age of the pulsar, so that they are expected to reflect the current spindown power. 
Thus, for the VHE domain some correlation
with $I\Omega\dot{\Omega}$ may be expected, with the strength of the correlation
depending on the magnitude of $\gamma_{\rm rad}$.

In the X-ray domain we observe only synchrotron radiation from freshly ejected electrons
(i.e. $\gamma \gg \gamma_{\rm rad}$), in which case we expect a relatively tight correlation
with the current spindown power. In fact, \cite{Che04} isolated the PWN X-ray contribution and found a
correlation between X-rays and spindown power of the form $L_X\propto (I\Omega\dot{\Omega})^{1.4\pm 0.1}$.

Recently \citet{hw03} derived birth periods for normal pulsars in the range 
$P_0\sim 0.6$ to 2.6 ms. For such birth periods this study would predict
very bright PWN at a distance of 10 kpc -- even by {\it EGRET} standards,
unless $\eta \ll 1$ during early epochs after birth. However, observations of the
\objectname[M1]{Crab nebula} does not support a low value of $\eta$ at birth.


Mature (Vela-like) PWN appear to have field strengths below $5\mu$G. If 
we also require the spectral turnover for such PWN due to radiative losses to be above 10 GeV,
the corresponding age of the PWN should be $\lesssim 300$ kyr as derived from Eqn. \ref{10gev}. 
Assuming that mature PWN evolve to such low field strengths in general, we infer that $\sim 50$ PWN
should be detectable for {\it GLAST/LAT} if we assume a birth period $P_0\le 0.5P$, whereas this
number reduces to 30 for $P_0\le 0.75P$. More may be visible if a significant amount of pulsars
are off-beam, in which case we may not see any pulsed counterpart. Another implication of an evolved PWN
with age near 300 kyr is that the current spindown power and magnetic field strength are too low
for a detectable synchrotron nebula, whereas the relatively low energy relic lepton component would still inverse 
Compton scatter external radiation fields into the GeV domain, which would lead to a population of ``dark high energy $\gamma$-ray sources'', i.e. those without multiwavelength counterparts.

To avoid source confusion and claim sources with confidence it would be important
to follow the {\it LAT} spectrum into the VHE $\gamma$-ray domain. The advantages for such
an association would be the following: (a) We detect the spectral turnover associated with $\gamma_{\rm rad}$,
(b) the positional agreement of the images between GeV and and the higher resolution TeV energies
makes the claim more significant, and (c) the size is expected
to shrink for photon energies $\gg 3kT\gamma_{\rm rad}^2$, with the image centroid converging
towards the associated pulsar as observed from \objectname{HESS\,J1825-137} \citep{Aha06b}.
Finally, to improve the accuracy in determining $P_0$, one should also add multiwavelength spectral information,
if observable.




\acknowledgments
The author would like to thank Mallory Roberts, Arache Djannati-Ata\"i 
and several members of the H.E.S.S. working group for supernova remnants, pulsars and plerions
for useful discussions on the VHE $\gamma$-ray and multiwavelength properties of plerions.

\end{document}